\documentclass[12pt,preprint]{aastex}
\usepackage{rotating}
\usepackage{CJK}

\newcommand{\tabincell}[2]{\begin{tabular}{@{}#1@{}}#2\end{tabular}}

\slugcomment{}
\shortauthors{Zhu et al.}
\begin{document}

\title{Predictions for Microlensing Planetary Events from Core Accretion Theory}
\begin{CJK*}{UTF8}{gkai}

\author{Wei Zhu (祝伟)\altaffilmark{1,2,3}, Matthew Penny\altaffilmark{2,4}, Shude Mao\altaffilmark{1,4}, Andrew Gould\altaffilmark{2}, and Rieul Gendron\altaffilmark{4}}
%\author{Wei Zhu \altaffilmark{1,2,3}, Matthew Penny\altaffilmark{2,4}, Shude Mao\altaffilmark{1,4}, Andrew Gould\altaffilmark{2}, and Rieul Gendron\altaffilmark{4}}

\email{weizhu@astronomy.ohio-state.edu}

\altaffiltext{1}{National Astronomical Observatories, Chinese Academy of Sciences, 20A Datun Road, Chaoyang District, Beijing 100012, China}
\altaffiltext{2}{Department of Astronomy, The Ohio State University, 140 W. 18th Ave., Columbus, OH 43210, USA}
\altaffiltext{3}{Department of Astronomy, Peking University, Yi He Yuan Lu 5, Hai Dian District, Beijing 100871, China}
\altaffiltext{4}{Jodrell Bank Centre for Astrophysics, University of Manchester, Alan Turing Building, Manchester M13 9PL, UK}

\begin{abstract}
  We conduct the first microlensing simulation in the context of planet formation model. 
  The planet population is taken from the Ida \& Lin core accretion model for $0.3M_\odot$ stars. 
  With $6690$ microlensing events, we find for a simplified Korea Microlensing Telescopes Network (KMTNet) the fraction of planetary events is $2.9\%$ , out of which $5.5\%$ show multiple-planet signatures. 
  The number of super-Earths, super-Neptunes and super-Jupiters detected are expected to be almost equal. 
  Our simulation shows that high-magnification events and massive planets are favored by planet detections, which is consistent with previous expectation. 
  However, we notice that extremely high-magnification events are less sensitive to planets, which is possibly because the 10 min sampling of KMTNet is not intensive enough to capture the subtle anomalies that occur near the peak. This suggests that while KMTNet observations can be systematically analyzed without reference to any follow-up data, follow-up observations will be essential in extracting the full science potential of very high-magnification events.
  The uniformly high-cadence observations expected for KMTNet also result in $\sim 55\%$ of all detected planets being non-caustic-crossing, and more low-mass planets even down to Mars-mass being detected via planetary caustics. 
  We also find that the distributions of orbital inclinations and planet mass ratios in multiple-planet events agree with the intrinsic distributions. 
\end{abstract}


%%%%%%%%%%%%%%%%%%%%%%%%%%%%
\section{Introduction}

Gravitational microlensing has discovered more than 50 extrasolar planets, although only about half (27) of them have appeared in the literature. Though few in number, microlensing planets occupy a unique region in the parameter space that is difficult to probe by other techniques \citep[e.g.,][]{gaudi2012,mao2012}. In fact, microlensing planets have already yielded interesting statistical results concerning the frequency of planets around M dwarf stars \citep{gould2010,cassan2012}, and intriguing possibilities of free-floating planets \citep{sumi2011} that can only be found by microlensing.

Ever since their first discovery, extrasolar planet systems have challenged the two fashionable models of planet formation - the core accretion and gravitational instability scenarios. In particular, in the core accretion theory, the planet population synthesis models are becoming increasingly sophisticated \citep[e.g.,][]{ida2010},  which take into account effects such as planetesimal accretion, gas accretion, disk evolution, migration and planet-planet interactions etc. For this reason, in this paper, we shall focus on this theory since its predictions are more quantitative and testable.

A detailed comparison between planet formation theory and microlensing observations has now become more imperative because of the emergence of next-generation microlensing experiments. In the past, the discovery of extrasolar planets often rested on a combination of work by survey teams [Optical Gravitational Lensing Experiment \citep[OGLE,][]{udalski2003} and Microlensing Observations in Astrophysics \citep[MOA,][]{bond2001}], and follow-up networks with higher-cadence observations [e.g., the Microlensing Follow-up Network \citep[$\mu$FUN,][]{gould2006,gaudi2008}, the Probing Lensing ANomalies NETwork \citep[PLANET,][]{albrow1998}, and RoboNet \citep{tsapras2009}]. However, these joint operations make the selection function sometimes difficult to quantify, although for some subsamples, such as the small number of very high-magnification events, the sample appears to be complete statistically \citep{gould2010}. Such a situation is likely to be changed significantly with the completion of the Korea Microlensing Telescopes Network (KMTNet) by the end of 2014. KMTNet will have three telescopes sited in Chile, Australia and South Africa \citep{kim2010}. Each telescope will have an aperture of 1.6m with 4 deg$^2$ of field of view, surveying about 4 fields with 10 minute cadence. With such high-cadence observations, KMTNet will be able to analyze the data without reference to any follow-up observations. Therefore the selection function will be much simpler. As a result, statistical results will be easier to obtain, which will provide a more robust measurements of planet abundances and distributions, which in turn will better constrain the planet formation theories.

\citet{shvartzvald2012} performed a detailed simulation for a next-generation microlensing network of 4 telescopes with aperture ranging from 1m to 1.8m and cadence from 15min to 45min. They use scaled Solar system analogs as the lens systems, and conclude that such a network can find of order 50 planets in 4 years, of which one in six reveals two planets in a single lensing event. This is a factor of several increase in discovery rate over the original alert/follow-up surveys \citep{gaudi2012}. KMTNet, with its larger field of view and higher cadence, is therefore expected to yield more planet detections and bring down the detection limit to lower-mass planets.

The present work is the first one that introduces the planet population synthesis model into microlensing simulations. Unlike previous works that use systems with only one planet, or simplified Solar system analogs as the lens system, our simulation is performed fully in the context of the Ida \& Lin core accretion model. Our results will be presented in two papers. The purpose of this paper is to provide a simple and yet somewhat realistic assessment of the fractions of extrasolar planets expected from KMTNet from the planet populations predicted by the core accretion theory; we will also explore how the KMTNet planet population differs from that of the current survey plus followup mode of discovery. In Paper II, we will focus on the multiple-planet events that are detected in our simulation; we will discuss the detection dependence of one planet on the other, the influence of the undetectable planets on the recovery of the parameters of detected planets, and the double/triple degeneracy \citep{gaudi1998,song2014}.

In \S2, we present the Ida \& Lin (2010) core accretion model, basics of microlensing and how we simulate microlensing data. In \S3 we describe our method of selecting events with extrasolar planets. In \S4, we present our main statistical results on the expected detection rates of extrasolar planet populations, and finally in \S5, we discuss further our results and implications for future observations.

%%%%%%%%%%%%%%%%%%%%
\section{Simulation Ingredients}
\subsection{Ida \& Lin core accretion model}

The planetary systems placed around our lenses are drawn from the \citet{ida2004a,ida2004b,ida2005,ida2008a,ida2008b,ida2010} core accretion planet population synthesis model. Their model generates protoplanetary disks with various surface densities and depletion timescales, based on observational constraints. Protoplanetary seeds are randomly selected in each disk, and integrated to protoplanets by accreting planetesimals, which are assumed to be formed from dust grains in the disk. Upon reaching a threshold mass, protoplanets begin to accrete the gas around them. Type I and Type II migrations and interactions between planets are also included to determine the final positions of the planets.

To compare with microlensing observations and simplify our computations, we place each system around a $0.3 M_\odot$ star, which is the most likely lens mass of microlensing \citep{gould1992,gaudi2008}. To avoid excessive calculations, we include only planets more massive than $0.1 M_\oplus$. We finally extract $669$ planetary systems from $1000$ Ida \& Lin systems.
We assume that all the planets in each system are coplanar, and randomly place their positions within the orbital plane according to their semi-major axes and eccentricities, and then choose a random isotropic orientation and project the system onto the sky.

%%%%%%%%%
\subsection{Microlensing}

Microlensing is most sensitive to planets that are close to the angular Einstein ring radius of the host lens \citep{mp1991,gould1992},
\[ \theta_{\rm E} = \sqrt{\kappa M_{\rm L} \pi_{\rm rel}};\ \kappa \equiv \frac{4G}{c^2 {\rm AU}} = 8.14 \frac{\rm mas}{M_\odot} . \]
Here $M_{\rm L}$ is the mass of the host lens, $\pi_{\rm rel}={\rm AU}(D_{\rm L}^{-1}-D_{\rm S}^{-1})$ is the lens source relative parallax, and $D_{\rm L}$ and $D_{\rm S}$ are the distances to the lens and source respectively. The typical timescale of a microlensing event, the Einstein radius crossing time, is
\[ t_{\rm E} = \frac{\theta_{\rm E}}{\mu_{\rm rel}} ,\]
where $\mu_{\rm rel}$ is the relative proper motion between the source and lens.

The morphology of microlensing light curves are strongly influenced by the caustic structures on the source plane. The caustic curve refers to the set of points on the source plane where the magnification of a point-like source is infinite. Caustics in the planetary microlensing case can be divided into three subclasses: (1) central caustic, referring to the caustic that is small and located close to the central star; (2) planetary caustic, referring to the caustic located far away from the star; and (3) resonant caustic, referring to a large but relatively weak caustic close to the central star when the planetary lens is close to the Einstein ring $\theta_{\rm E}$.

The Ida \& Lin systems are normally multiple-planet systems. Therefore we use the multiple-lens microlensing theory in our simulation. The multiple-lens equation can be written as \citep{witt1990}
\begin{equation} \label{eq:lens-equation}
	z_s = z-\sum_{k=1}^N \frac{q_k}{\bar{z}-\bar{z}_k},
\end{equation}
where $z_s$ is the complex position of the source, $q_k$ is the mass ratio of the $k^{\rm th}$ lens relative to the mass of the massive lens, in our case, the central star $M_{\rm L}$, $\bar{z}_k$ is the conjugate position of the $k^{\rm th}$ lens, and $z$ and $\bar{z}$ are the positions of the corresponding image in complex and conjugate form respectively. Distances in Equation (\ref{eq:lens-equation}) are scaled by the Einstein radius of the host star at the lens plane,
\[ R_{\rm E} = D_{\rm L} \theta_{\rm E} .\]

For simplicity we treat the lens system as static (i.e., no orbital motion) and ignore the microlens parallax. The ray-shooting method \citep{schneider1986,schneider1987} is used to generate theoretical light curves for multiple-lens microlensing. In reality, the timescale of microlensing events, $\theta_{\rm E}$, may vary from event to event, but to avoid complicated ray-shooting process and also to get comprehensive statistical results in a reasonable computation time, we fix $t_{\rm E}$ to 15.7 days, which is for a typical bulge microlensing event with the lens system at $D_{\rm L}=7.4$ kpc, the source star at $D_{\rm S}=8.6$ kpc, the lens star with mass $M_{\rm L}=0.3M_\odot$ and the relative proper motion between the source and the lens $\mu_{\rm rel}=5$ mas/yr. These also yield $R_{\rm E}=1.59$ AU. Additionally, we adopt a typical turnoff source star with radius $R_\star=1.6R_\odot$, which corresponds to a scaled source size $\rho=\theta_\star/\theta_{\rm E}=0.004$; a uniform surface brightness profile (i.e., no limb darkening effect) is also employed. The density of rays used in the ray-shooting program is determined according to \citet{dong2006}. Based on the above settings, we finally shoot $4.9\times10^9$ rays over the area from (-4,-4) to (4,4) to obtain an averaged accuracy of $\sim 0.1\%$ when the source is unmagnified. This calculation error is much smaller than the simulated photometric error (see below), so we do not account for it in $\chi^2$.

\begin{figure}
\centering
\plotone{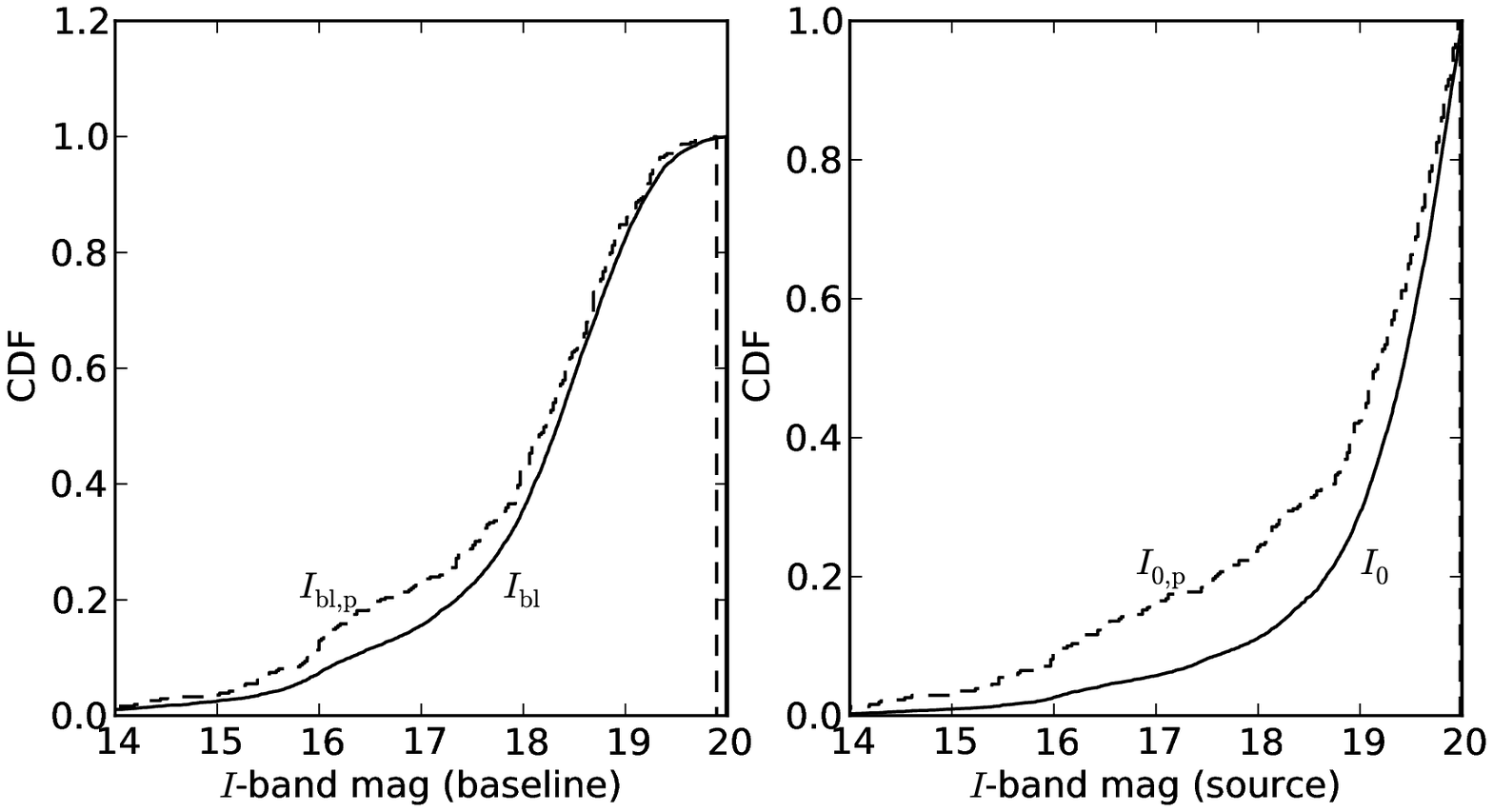}
\caption{Cumulative distributions of baseline magnitudes of all simulated events (solid line) and planetary events (dashed line). \textit{Left panel:} the base $I$-band magnitudes; \textit{right panel:} the base $I$-band magnitude for the source.
\label{fig:magnitudes}}
\end{figure}

The impact parameter $u_0$ for each event is randomly chosen from $-0.3$ to $0.3$, meaning that the maximum magnification of each event is above 3.4. We do this to ensure an interesting number of planet detections for the available computing time, and a reasonable number of high-magnification events. Our simulation covers the time $-1.5t_{\rm E} \le t-t_0 \le 1.5t_{\rm E}$, where $t_0$ is the time of closest approach relative to the host star in the lens system. The total baseline $I$-band magnitudes $I_{\rm bl}$ and blending fractions $f_{\rm bl}$ are drawn from \citet{smith2007} where the blending effects of typical Galactic bulge fields with high stellar density are simulated for OGLE. The cumulative distributions of $I_{\rm bl}$ and the source baseline magnitude $I_0$, which is related to $f_{\rm bl}$ by $I_0=I_{\rm bl}-2.5\log_{10}{f_{\rm bl}}$, are displayed in Figure~\ref{fig:magnitudes}. The photometry is simulated using the same code as \citet{penny2011}, but employing parameters that better match the KMTNet survey \citep{kim2010}. Specially, we assume a larger telescope diameter of 1.6 m, a worse mean seeing of 1.4 arcsec and a larger systematic error floor of $0.5\%$. The chosen seeing is therefore worse than that used to estimate the blending statistics, so the amount of blending is slightly underestimated.

Each planetary system is used to generate 10 light curves, so we end up with 6690 microlensing events in total.

%%%%%%%%%%%%%%%%%%
\section{``Observing'' Simulated Data}

We employ the following procedures to determine how many planets are detected in each microlensing event. Simulated light curves are first fitted with a single-lens model with six parameters $t_0$, $t_E$, $u_0$, $I_{\rm bl}$, $f_{\rm bl}$, and $\rho$. We minimize $\chi^2$ using the \texttt{MINUIT} routine from \texttt{CERNLIB} \citep{james1975}, with the first three parameters free and the last three parameters fixed to the true value to simplify the calculations. This is conservative in the sense that allowing the last three parameters to vary can only increase $\Delta \chi^2$, and thus improve the derived robustness of the detection. If the difference in $\chi^2$ between the best-fit single-lens model and the theoretical model used to simulate the light curve, $\Delta \chi_{\rm single}^2$, is larger than 200, this event is considered a potential planetary event. From $6690$ microlensing events, we find $313$ events with $\Delta \chi_{\rm single}^2>200$.

If a light curve is assessed as possibly containing planetary perturbations, we employ the following method to identify which planets are responsible for these perturbations.
We first rank the planets in the system in decreasing order by the width of their planetary caustics \citep{han2006}
\begin{equation}
w_k = 
\left\{ \begin{array}{ll}
	q_k^{1/2} |z_k|^3 & {\rm if} |z_k|<1 \\
	q_k^{1/2} |z_k|^{-2} & {\rm if} |z_k| \ge 1 
	\end{array} \right.
\end{equation}
Theoretical double-lens (i.e., the host star plus a single planet) light curves are then generated for the six most highly ranked planets individually, using the same $u_0$, $I_{\rm bl}$, $f_{\rm bl}$ and under the same simulated photometric conditions. Such double-lens light curves are also fitted with a single-lens model. 

\begin{figure}
\epsscale{0.8}
\centering
\plotone{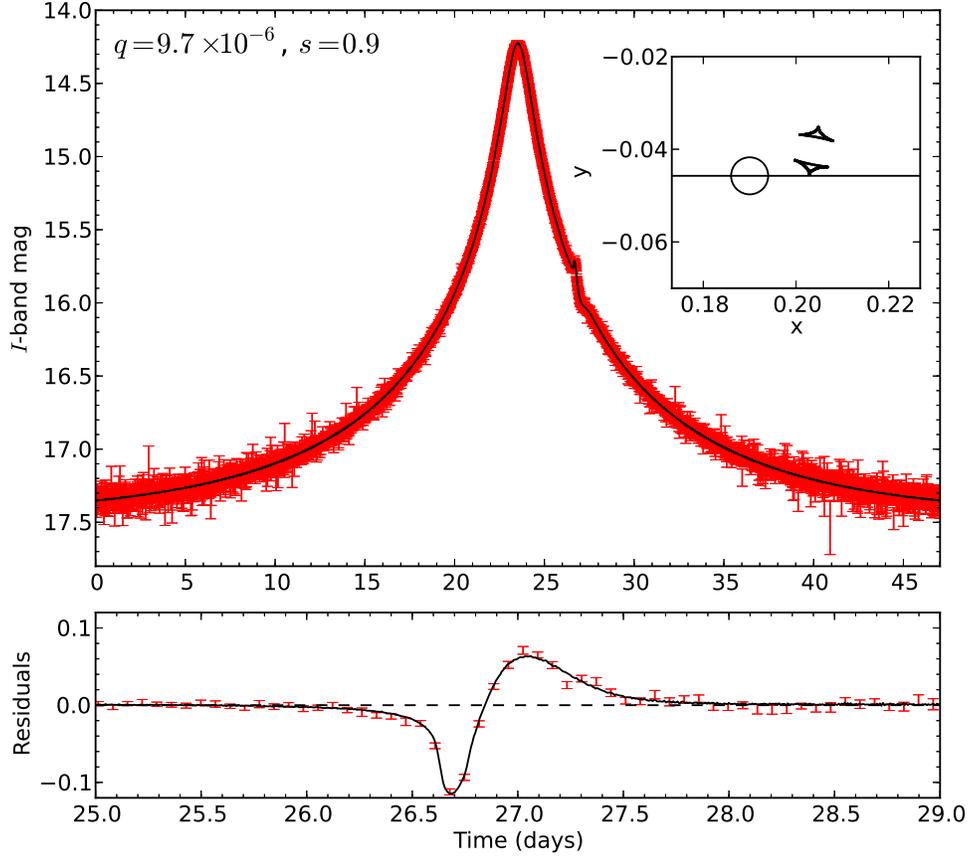}
%\plottwo{lc-1991.eps}{lc-1991-color.eps}
\caption{\textit{Top}: The simulated light curve and the best-fit single-planet model for No. 1991 event in our simulation. \textit{Bottom}: Difference between this model and the best-fit single-lens model; the data are re-binned with every 10 points in this panel. The mass ratio $q$ and separation $s$ of the planet is indicated, which correspond to a mass of $\sim 1M_\oplus$ and physical separation of $\sim1.4$ AU.
\textit{Inset}: Source path through the caustic geometry; the source size $\rho$ is indicated. 
%See the electronic edition of the Journal for a color version of this figure.
\label{fig:lc-single}}
\end{figure}

We examined by eye each of the 313 candidates of planetary event generated by the automatic selection criterion. We rejected 21 of these as not real detections. Of these 19 were events with very bright sources and hence extremely small error bars, for which the numerical precision of our light curve modeling (designed for more typical events) was not adequate. Hence, the $\Delta \chi^2$ was simply due to numerical noise. For the other two, the $\Delta \chi^2$ was contributed from more than one planets, but a single-planet model yielded $\Delta \chi^2<200$, and the signal-to-noise ratio was not adequate to claim more than one planet. Such a ``planet detection'' would be rejected in practice and so these two events were excluded. Finally we confirm $292$ planetary events, of which $23$ have two detectable planets and none has more than two. In Figure~\ref{fig:lc-single} we show an example single-planet event which contains an Earth-mass planet.

That two planets can be detected individually does not guarantee that they can be detected in the same event, because of the degeneracy existing between light curves arising from multiple- and single-planet events \citep[e.g.,][]{gaudi1998,song2014}. Thus a double-lens model is used to fit the $23$ multiple-planet candidates. We first fit each light curve without parallax and orbital motion. For some of them we tried with parallax and orbital motion included, but did not see significant improvement in $\chi^2$ (recall that the events are simulated assuming no such effects). 

\begin{figure}
\epsscale{0.8}
\centering
\plotone{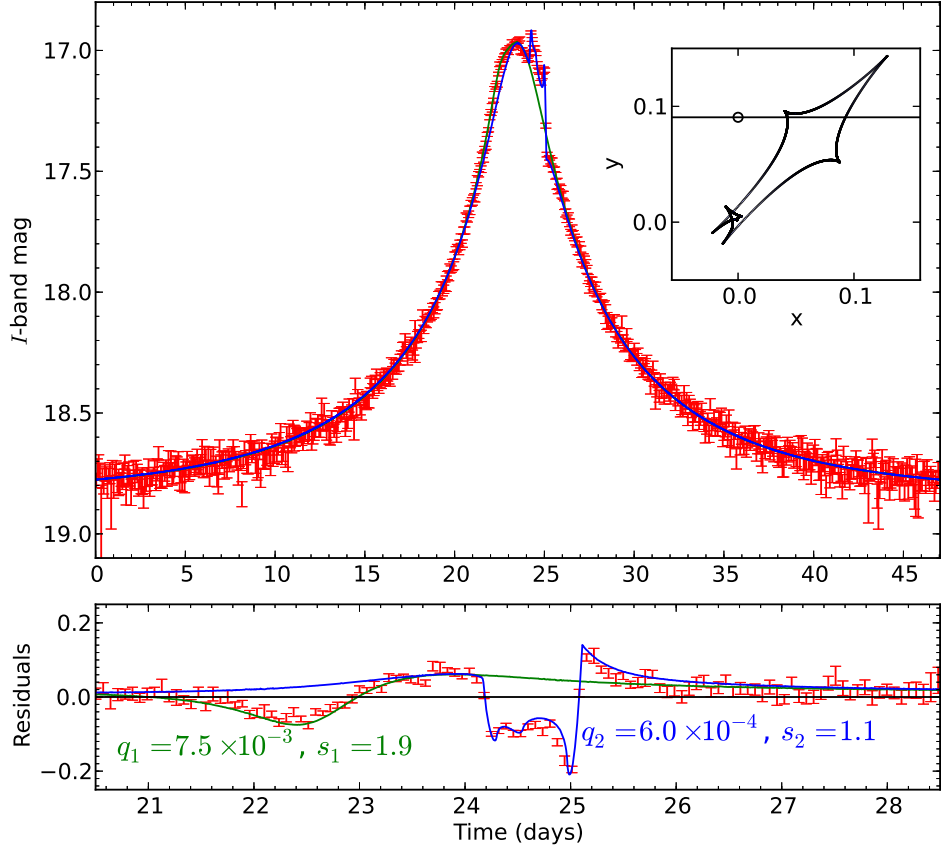}
%\plottwo{lc-3451.eps}{lc-3451-color.eps}
\caption{\textit{Top}: The simulated light curve and two single-planet light curves for No. 3451 event in our simulation. \textit{Bottom}: Difference between the best-fit single-lens model and the data (red points with error bars), the single-planet light curve with Planet 1 (green), and the single-planet light curve with Planet 2 (blue); the planetary signature can be explained by a combination of these two planets. In both panels the data are re-binned with every 10 points. Planet 1 has mass $\sim2.2M_{\rm J}$ and separation $\sim 3$ AU, and Planet 2 has mass $\sim 60M_\oplus$ and separation $\sim1.8$ AU. \textit{Inset}: Source path through the caustic geometry; the source size $\rho$ is indicated. 
%See the electronic edition of the Journal for a color version of this figure.
\label{fig:lc-double}}
\end{figure}

The confirmation of multiple-planet events is very subtle since few related studies have been done based on real data. However it is generally believed that the confirmation of a second planet requires a higher $\Delta \chi^2$ threshold because, with a much lower occurrence of multiple-planet events compared to that of single-planet events, the small deviations in the light curve after substraction of the first planet have a larger probability to be explained by systematics or other stellar variabilities \citep[e.g.,][]{gould2013}. We use $\Delta \chi^2_{\rm double}$ to denote the $\chi^2$ difference between the best-fit single-planet (double-lens) model and the theoretical model. 
Of the 23 multiple-planet candidates, 4 have separate planetary signatures caused by different planets, so they are very secure detections. Among the remaining 19 candidate events, we find 4 events with $\Delta \chi^2_{\rm double}<200$, 3 with $200<\Delta \chi^2_{\rm double}<300$ (all below 250), 6 with $300<\Delta \chi^2_{\rm double}<400$ (only one below 350), and 6 with $\Delta \chi^2_{\rm double}>400$. The event with $\Delta \chi^2_{\rm double}$ between 300 and 350 are confirmed by our experienced microlensing observer (A.G.), while the 3 with $\Delta \chi^2_{\rm double}$ between 200 and 250 are not, although one of them seems to show plausible signatures. Therefore, we set the $\Delta \chi^2$ threshold to be 300, with which we finally confirm 16 two-planet events. 
A more detailed discussion about these two-planet events will be presented in Paper II. An example double-planet event is shown in Figure~\ref{fig:lc-double}.

%%%%%%%%%%%%%%%%%%%%%%%%%%%
\section{Results}
\subsection{Dependence on mass, separation and caustic type} \label{subsec:q-s-dependence}

Our simulation results in 308 detected planets, including 276 from single-planet events and 32 from double-planet events, from 74560 planets in 6690 lens systems, each with at least one planet. The baseline magnitudes of these planetary events are shown in cumulative function form in Figure~\ref{fig:magnitudes}, together with that of all our simulated microlensing events for comparison. 

Masses and separations of all detected planets are shown in the lower panel of Figure~\ref{fig:q-s-map}, with the histogram of the separations shown on the top. Here we use the true masses and separations not the mass ratios and separations obtained by fitting the light curves. Although the latter is what we should use in order to compare with real observations, we notice that $85\%$ of all planetary events can be very well reproduced by the single-planet light curve derived from the true mass and separation. Colors and symbols in Figure~\ref{fig:q-s-map} encode the caustic that was encountered and the number of detected planets, respectively. As is expected, planets are mostly detected near $R_{\rm E}$, although high-mass planets ($M_p>M_{\rm Neptune}$, and hereafter) can be detected in a broader range than the low-mass planets. We also find that low-mass planets are more often detected via their planetary caustics, while high-mass planets are more often via central caustics. The planets detected via resonant caustics are located within a more narrow region around $R_{\rm E}$, compared to the overall distribution. We notice that all the sub-Earths ($M_p<M_\oplus$) are detected via planetary caustics, which is reasonable, since for fixed $s$, the diameter of the planetary caustic scales with $\sqrt{q}$ while the diameter of the central caustic scales with $q$. In particular, we notice that even Mars-mass planets can be detected in this KMT-like simulation. Figure~\ref{fig:q-s-map} also shows that planets in double-planet events are mostly high-mass and detected via central caustics. We will give our explanation of this in Section~\ref{sec:discussion}.

\begin{figure}
\centering
\plotone{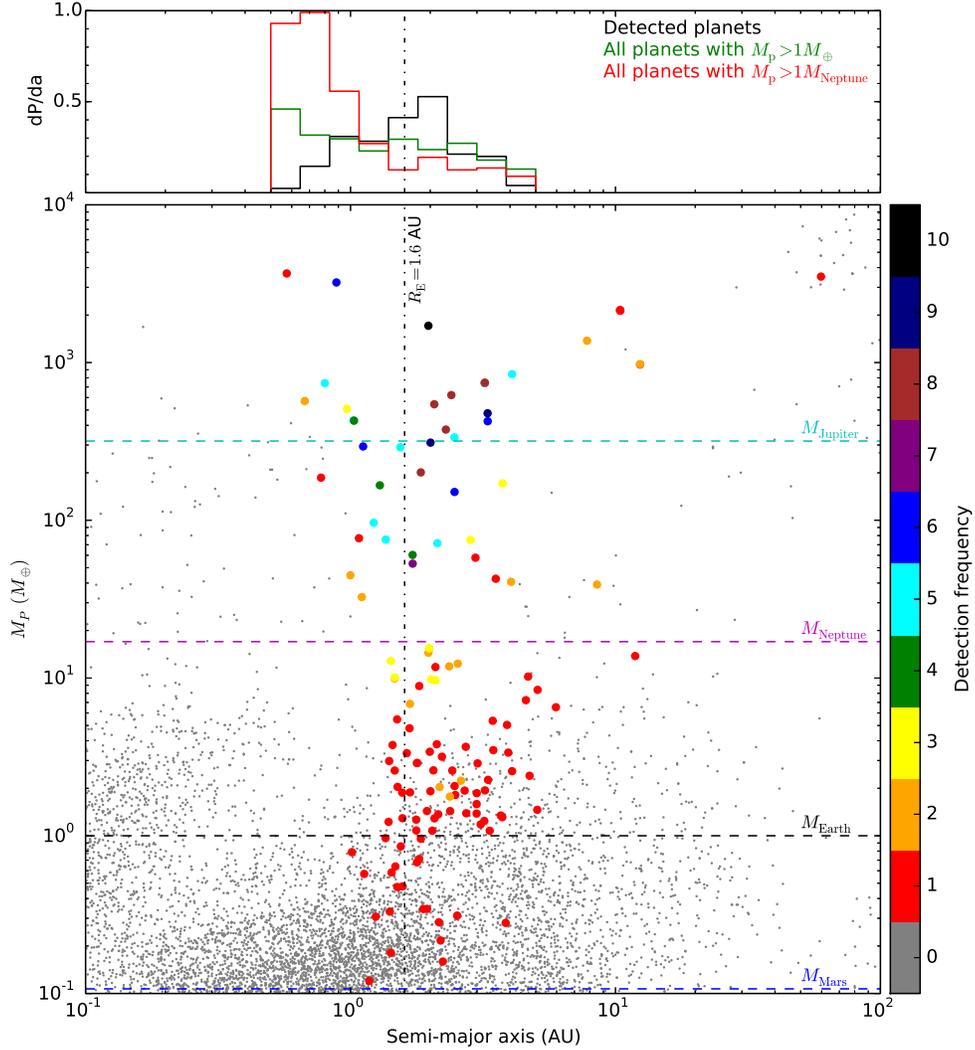}
\caption{\textit{Lower panel}: the distribution of all Ida \& Lin planets in the mass vs. semi-major axis plane; colors encode the detection frequency of the planet within 10 simulated events using the same planetary system. \textit{Upper panel}: the (normalized) histogram of the semi-major axes of detected planets (black) with the detection frequency considered, all planets with $M_{\rm p}>1M_\oplus$ (green) and with $M_{\rm p}>1M_{\rm Neptune}$ within the same semi-major axis range. The vertical line indicates the position of the Einstein ring radius in the lens plane.
\label{fig:mass-sma}}
\end{figure}

We show the distribution of masses and semi-major axes of Ida \& Lin planets in Figure~\ref{fig:mass-sma}, with colors in the lower panel encoding the detection frequency within 10 simulated events based on the same planetary system. As has been seen in Figure~\ref{fig:q-s-map}, the high-mass planets are more often detected than those of low mass. 
As expected, the distribution of semi-major axes $a$ is shifted upward relative to the Einstein radius by $\Delta \log{a} \sim 0.5\log{1.5}=0.09$, while the distribution of $\log{a_\perp}$ ($a_\perp$ is the projected separation $s$) in Figure~\ref{fig:q-s-map} peaks right at the Einstein ring. What is more surprising is that both distributions are approximately symmetric even though the underlying distribution of Ida \& Lin planets rises strongly toward closer separations, which means that wide planets have larger detection efficiencies than close ones. Such a detection bias needs further study in order to recover the underlying distribution of planets based on real microlensing observations.

%%%%%%%%%%%%%%%%%%%%%%%%
\subsection{Dependence on impact parameter or maximum magnification} \label{subsec:u0-dependence}

We correct the impact parameter to the center of magnification ($u_0^\star$), rather than that to the host star ($u_0$) which was randomly chosen in the simulation, 
\begin{equation}
 u_0^\star = u_0-\sum_k \frac{q_k \sin{\alpha_k}}{|z_k|+|z_k|^{-1}} ,
\end{equation}
where $\alpha_k$ is the angle between the source trajectory and the $k^{\rm th}$ planet \citep{chung2005}.
Figure~\ref{fig:u0} shows the cumulative distribution function of this impact parameter $u_0^\star$ for different groups of events in our simulation. Compared to the uniformly distributed input impact parameters, relatively small impact parameters, which correspond to relatively high-magnification events ($A_{\rm max} \approx 1/u_0^\star$), are favored in planet detections, especially multiple-planet detections, which is consistent with previous theoretical expectations \citep{griest1998}.

However, we notice that events with extremely small impact parameters ($u_0^\star \lesssim 0.005$) are less sensitive to planets than moderately small $u_o^\star$.
To clarify this, we list the number of events within each $A_{\rm max}$ range in Table~\ref{tab:Amax-number}. Events with $A_{\rm max}>200$, which can be regarded as extremely high-magnification events in our simulation due to the relatively large source size we use, are significantly less sensitive to planets than the lower-magnification ones. This seems to conflict with previous studies based on ongoing observations \citep{gould2010}. The reason might come from the different observing strategies used in current observations and in our simulation. We will give a detailed discussion on this in Section~\ref{sec:discussion}.

%%%%%%%%%%%%%%%%%%%%%%%
\subsection{Dependence on inclination and planet mass ratios} \label{subsec:i0-q}

For microlensing planets, the orbital inclination is important in converting the projected separation to semi-major axis of the planet and therefore understanding the physical properties of planetary systems. Thus the distribution of orbital inclinations of planetary events in realistic simulations are worthy of investigation.

\begin{figure}
\centering
\plotone{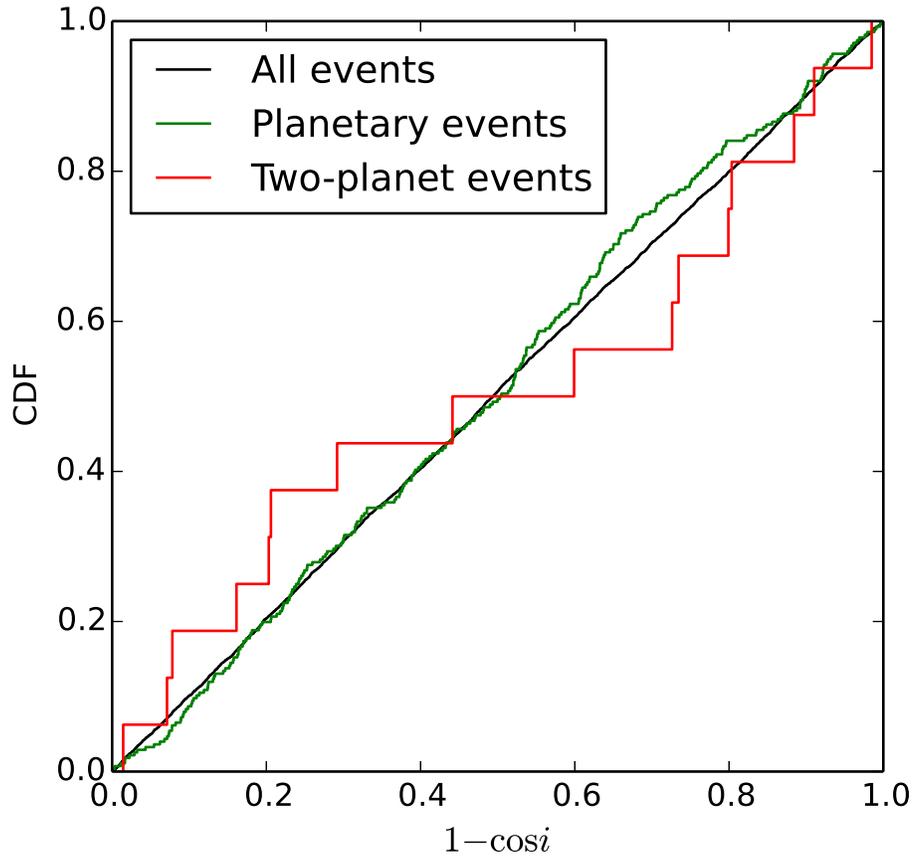}
%\plottwo{inclination.eps}{inclination-color.eps}
\caption{Cumulative distribution of orbital inclinations for three groups of events: all microlensing events (black), all planetary events (green), and double-planet events (red). From left to right a system goes from face-on to edge-on. 
%See the electronic edition of the Journal for a color version of this figure.
\label{fig:i0}}
\end{figure}

Our simulation shows that the distribution of orbital inclinations of detected planetary events are statistically consistent with the input distribution of inclinations, as Figure~\ref{fig:i0} shows, although the distribution for double-planet events deviates slightly from the input distribution. This implies that the planet detection is not biased on any typical orbital inclinations, which is reasonable since the projected position is determined not only by the inclination but also by the orbital phase of the planet.

\begin{figure}
\centering
\plotone{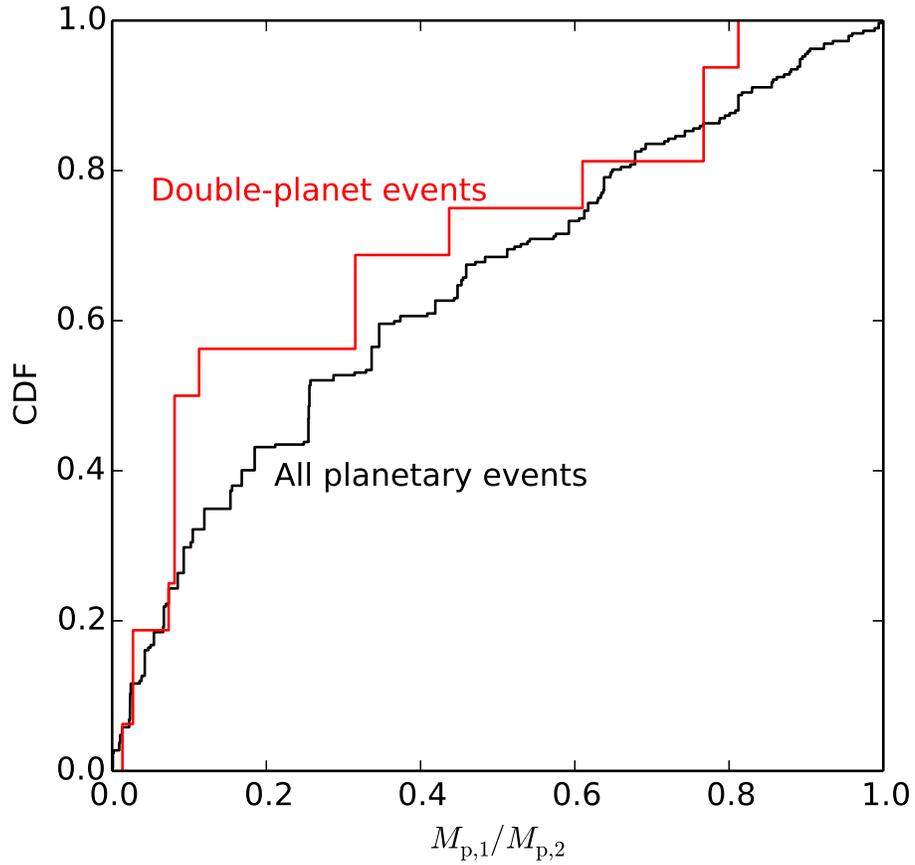}
%\plottwo{mass-ratio.eps}{mass-ratio-color.eps}
\caption{The cumulative distribution of mass ratios between planets in double-planet events (red) and that between two most massive planets, whether detected or not, for all planetary events (black). 
%See the electronic edition of the Journal for a color version.
\label{fig:mass-ratio}}
\end{figure}

Figure~\ref{fig:mass-ratio} shows the comparison between the mass ratios of planets detected in double-planet events and that of the two most massive planets in all planetary events. A two-sample KS test gives a confidence level $\alpha=34\%$, meaning the two distributions are consistent with being drawn from the same distribution. This indicates that multiple-planet systems detected via microlensing are representative of all multiple-planet systems.

\subsection{Detection efficiency} \label{subsec:detection-efficiency}

\begin{figure}
\centering
\plotone{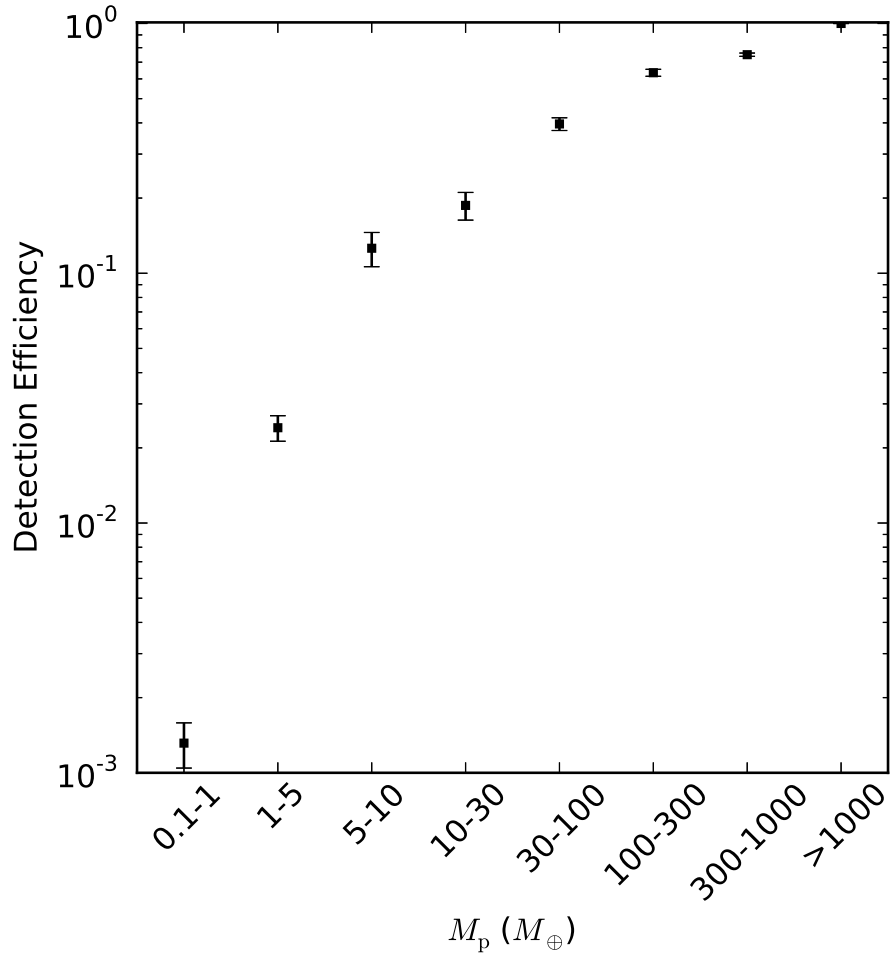}
\caption{Detection efficiency as a function of the mass of planets located between $0.5R_{\rm E}$ and $2R_{\rm E}$.This is the visualization of Table~\ref{tab:efficiency-mp} but in more bins; the error bars are the binomial statistical errors due to the number count.
\label{fig:efficiency-mp}}
\end{figure}

We estimate the detection efficiency of microlensing for planets located between $0.5R_{\rm E}$ and $2R_{\rm E}$. The numbers of planets with different masses and the fraction of detected planets are listed in Table~\ref{tab:efficiency-mp} and displayed in Figure~\ref{fig:efficiency-mp} in more bins with statistical error bars. In particular, we notice that the detection efficiency for super-Jupiters is 20 times higher than that for Earth-mass planets. However, the low detection efficiency of low-mass planets is compensated by the larger number of such planets, which yields almost equal numbers of super-Earths, super-Neptunes and super-Jupiters detected in our simulation. 
This is consistent with the result of \citet{henderson2014}, wherein a more realistic simulation for KMTNet is performed and the distribution of detected planets is estimated based on the planetary mass function given by \citet{cassan2012}.

As a result of this difference in detection efficiency, the fraction of massive planets detected in our simulation exceeds the fraction of such planets given by the planet formation model. Within $292$ planetary events, we find $114$ systems holding a super-Jupiter planet. This fraction, $39\%$, is $\sim 8$ times higher than the prediction of the Ida \& Lin model, which only contains $5.3\%$ of such systems.

%%%%%%%%%%%%%%%%%%%%%%%%%%
\section{Discussion} \label{sec:discussion}

We conducted a simple and yet realistic microlensing simulation for a KMTNet-like microlensing survey. The planet population is taken from the Ida \& Lin core accretion model for $0.3M_\odot$ lenses. Our simulation results in 292 planetary events, including 16 double-planet events, from 6690 microlensing events for which the lens system has at least one planet more massive than $0.1M_\oplus$. With the frequency of such planetary systems considered, we find the fraction of planetary events is $2.9\%$, out of which $5.5\%$ show multiple-planet detections. 

\begin{figure}
\plotone{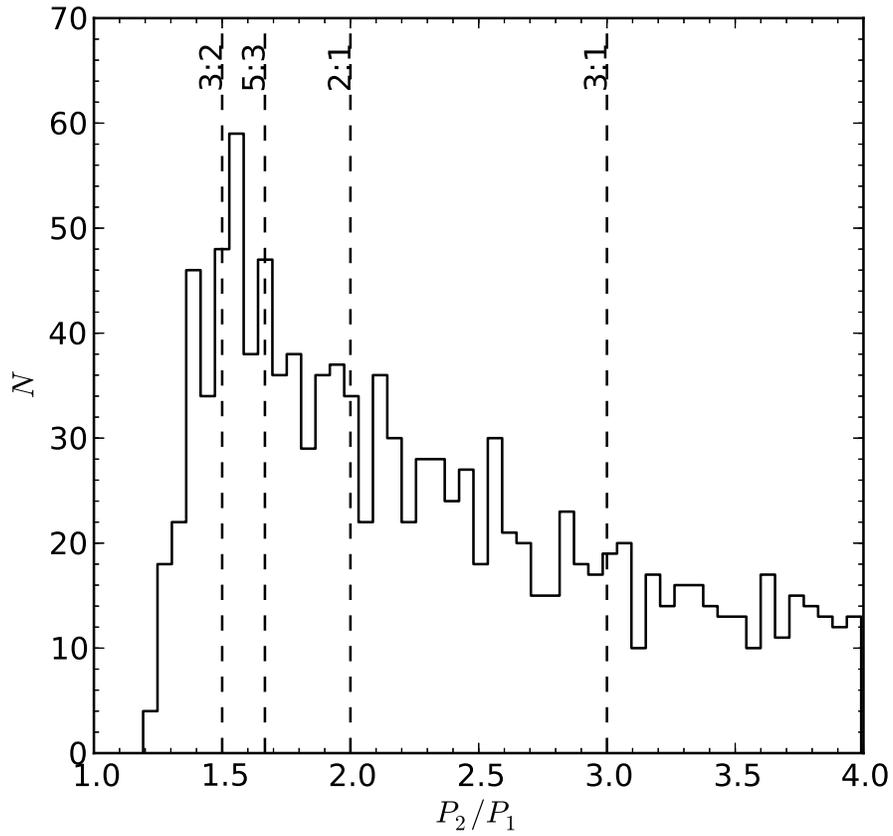}
\caption{The histogram of period ratio $P_2/P_1$ for all Ida \& Lin planet pairs with mass greater than $1M_\oplus$. The dashed lines indicate the positions of exact 3:2, 5:3, 2:1 and 3:1 resonances.
\label{fig:period-ratio}}
\end{figure}

We address the limitations of our simulation here before discussing the implications of our results.
(1) We admit that our simulation is not fully realistic in the sense that the Einstein timescale $t_{\rm E}$, which involves distances $D_{\rm S}$ and $D_{\rm L}$, lens mass $M_{\rm L}$ and relative proper motion $\mu_{\rm rel}$, and the source size $\rho$, are fixed to some typical values. This does prevent us from making precise predictions for KMTNet, but predicting the yields of KMTNet or any other specific microlensing experiment is not the main purpose of our work. Our simulation aims to address more general questions, which are not easily clarified if too many observational factors are considered. Moreover, a microlensing simulation with 12 lenses in each system on average will become extremely complicated if all the parameters that we have held fixed (i.e., $D_{\rm L}$, $D_{\rm S}$, $M_{\rm L}$, $\mu_{\rm rel}$ and $\rho$) are set free.
(2) The planet population given by Ida \& Lin's model is produced for stars in the Galactic disk, but in our simulation the lens system is placed in the Bulge. Planets forming around Bulge stars may well have very different distributions from those forming around Disk stars, not only because of the different metallicity but also because of the very dense environment \citep{thompson2013}. However, there is as yet no model available to quantitatively predict the planet population in the Galactic Bulge. Therefore, using planet population predictions for Disk stars is our only choice. One positive outcome of this approach is that comparing the results of our simulation with real observations may tell us how different the planet populations are in the Bulge and in the Disk, which is a question that can only be answered by microlensing.
(3) When randomly placing the planets on their orbit, we do not take mean motion resonances into account. This may not be correct in the case of resonant systems. However, we notice that the Ida \& Lin's model does not show strong resonance signatures, as is shown in Figure~\ref{fig:period-ratio}. Therefore, the orbit of each planet is mostly unaffected by others in the same system, so randomly placing them on their orbit is acceptable.

\begin{figure}
\centering
\plotone{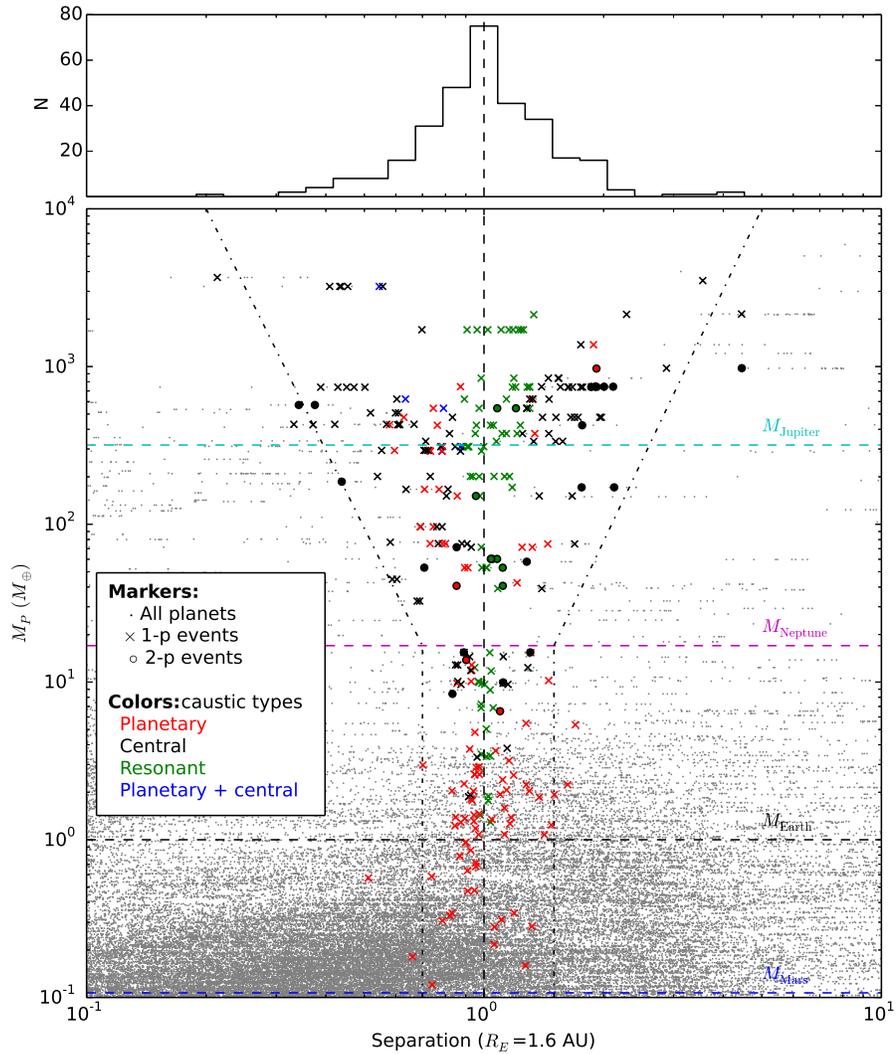}
\caption{\textit{Lower panel}: the distribution of all Ida \& Lin planets (grey), and detected planets in our simulation including single-planet microlensing events (crosses) and multiple-planet events (circles). Colors represent what kind of caustics this planet is detected with: red for planetary caustic, green for resonant caustic, black for central caustic, and blue for both planetary and central caustics involved. The dash-dotted lines indicate the rough boundary of these planets. \textit{Upper panel}: the histogram of the separations of all detected planets.
\label{fig:q-s-map}}
\end{figure}

\begin{figure}
\centering
\plotone{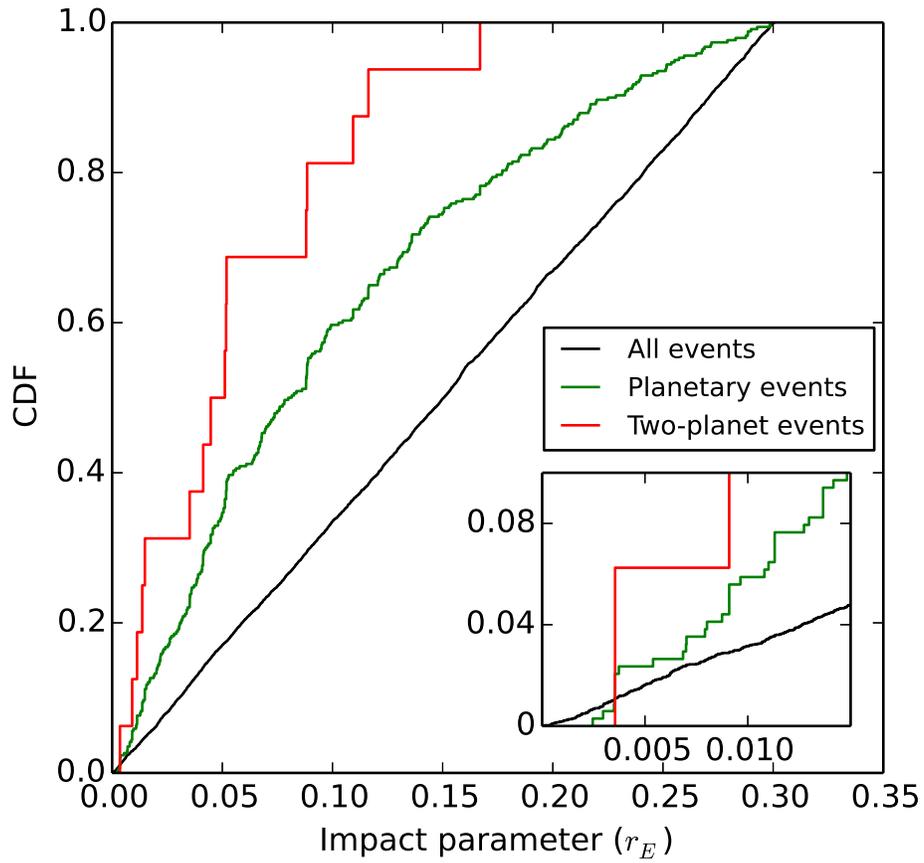}
%\plottwo{impact-parameter.eps}{impact-parameter-color.eps}
\caption{Cumulative distribution of impact parameters for three groups of events: all microlensing events (black), all planetary events (green), and double-planet events (red). 
%See the electronic edition of the Journal for a color version of this figure.
\label{fig:u0}}
\end{figure}

Our simulation yields more multiple-planet events than our naively expectation. Given that the total number of planets in $6690$ systems is $74560$, the probability for one planet to be detected is therefore $p=0.0041$ if we naively assume such detection does not depend on the characteristics of the microlensing event or properties of the planet. Then the number of single-planet events we would expect to detect in our simulation should be
\begin{equation}
 N_1 = \sum_j n_j p ,
\end{equation}
and the number of double-planet events is
\begin{equation}
N_2 = \sum_j \frac{n_j(n_j-1)}{2} p^2 ,
\end{equation}
where $n_j$ is the number of planets in the $j^{\rm th}$ system. Given $N_1=276$ in our simulation, we would expect $N_2$ should be $7$ if our assumption holds, which is significantly lower than what we do detect, $N_2=16$. This is reasonable since the detectability of planets depends not only on the physical properties of the planet, but also on the impact parameter of that microlensing event. High-magnification events and massive planets are more favored in multiple-planet microlensing, as is shown in Figures~\ref{fig:q-s-map} and \ref{fig:u0}. Our simulation therefore predicts that multiple-planet events will be detected more than our naive expectation, but they are strongly biased toward massive planets and higher-magnification events.

In Section~\ref{subsec:u0-dependence} we have shown that extremely high-magnification events are less sensitive to planet detections than those moderately high-magnification ones in such a KMTNet-like survey program. This apparently conflicts with previous theoretical predictions \citep{griest1998} as well as ongoing observations \citep{gould2010}. The reason might come from the different observing strategies used in our simulation and in current observations. The survey plus follow-up mode used in current microlensing observations can achieve very high cadence (e.g., more than one observation per minute after accounting for multiple observatories) during the peak of high-magnification events, although the survey teams typically obtain only a few observations per night. These intensive observations during the peak make high-magnification events extremely sensitive to planet perturbations. Our simulation is conducted using a strategy similar to that expected for KMTNet, which uses a constant cadence (10 mins) of observations everywhere. Therefore, for extremely high-magnification events, the planet perturbation is so weak that it may be missed by such a 10-min cadence observing strategy. 
This argues that even in the era of next generation surveys, there is still a need for follow-up of high-magnification events, which will require the next generation surveys to process their data in real time and produce high-magnification alerts, as is done for current surveys. With $\sim 20\%$ of high-magnification planet detections yielding multiple planets, such follow-ups are important for measuring the number of multiple-planet systems \citep{gaudi1998}.

The advantage of conducting uniformly high-cadence observations everywhere in the light curve like KMTNet, in addition to obtaining a well controlled planets sample for statistical studies, is the ability to detect more low-amplitude planetary perturbations and perturbations due to planetary caustics. Low-amplitude perturbations are usually produced by source-star trajectories that do not cross any caustics. If we define caustic crossings as occurring if the closest distance between the source trajectory and caustics is less than two source radii, we find that $55\%$ of all detected planets in our simulation are not due to this caustic crossing, as is listed in Table~\ref{tab:Amax-number}. 
In contrast, we searched all published microlensing planets and characterized them according to this definition of caustic crossing. We find only three real microlensing planets are due to such non-caustic-crossing events within $26$ published microlensing planets with very good data coverage under the current observation strategy which are listed in Table~\ref{tab:all-planets}. 
This implies that in future microlensing programs like KMTNet, WFIRST \citep[the Wide-Field InfraRed Survey Telescope,][]{spergel2013} and possibly Euclid \citep{penny2013}, at least half of the microlensing planets will not be detected by crossing caustics. The non-caustic-crossing character of the event makes it more difficult to determine the physical properties of the lens system, since the unknown but important quantity $\theta_{\rm E}$ cannot be determined from the angular size of the source star that is derived from the source color and brightness \citep{yoo2004}. However, in the case of WFIRST it may be possible to measure $\theta_{\rm E}$ by astrometric microlensing \citep{gould2014} or (in the case that the lens is luminous) by taking high-resolution images several years before or after the event.

The number of planets detected via planetary, central and resonant caustics are 107, 128 and 78 respectively.
\footnote{Planets detected via both planetary and central caustics are counted twice.}
 The fraction of that by planetary caustics, $35\%$, is slightly higher than but consistent with $27\%$(=7/26) based on real microlensing planets. More planets being detected via planetary caustics and the high cadence observations around the planetary anomaly lead to the detection of very low-mass planets even down to Mars-mass, as the planetary caustic shrinks slower ($\sim \sqrt{q}$) than the central caustic does ($\sim q$) as the planetary mass ratio $q$ decreases.

\begin{figure}
\centering
\epsscale{0.5}
\plotone{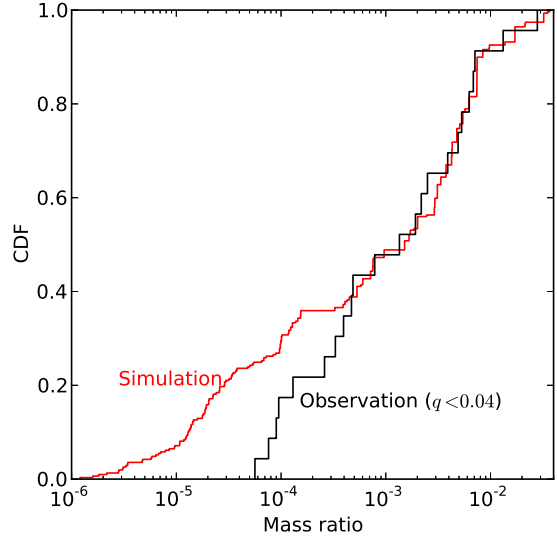}
%\plottwo{cdf-q.eps}{cdf-q-color.eps}
\caption{The cumulative distribution of mass ratio $q$ of detected planets in our simulation compared to real observations. We only include microlensing planets with $q<0.04$ since this is the upper limit of mass ratio in our simulation, which corresponds to a $13M_{\rm J}$ planet around a $0.3M_\odot$ star. 
%See the electronic edition of the Journal for a color version.
\label{fig:cdf-q}}
\end{figure}

\begin{figure}
\centering
\epsscale{0.8}
%\plotone{cdf-q-2sample.eps}
\plotone{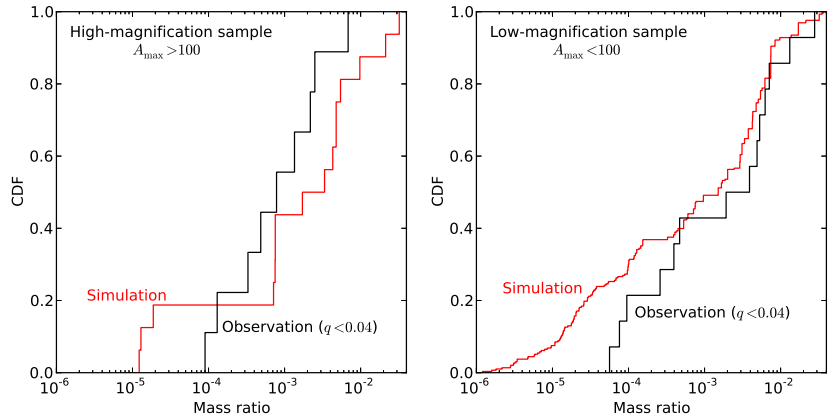}
\caption{Similar to Figure~\ref{fig:cdf-q}, instead that now the whole planet samples are divided into two groups: planets detected in high-magnification events (\textit{left panel}) and planets detected in low-magnification events (\textit{right panel}). 
%See the electronic edition of the Journal for a color version.
\label{fig:cdf-q-2sample}}
\end{figure}

In Figure~\ref{fig:cdf-q} we compare the cumulative distribution of mass ratios of planets detected in our simulation with that of real microlensing planets. we notice the two curves coincide with each other surprisingly well for $q>10^{-3}$, but that the curve from our simulation has a long tail toward very small mass ratio, which means that future microlensing surveys will be able to explore more very low-mass planets than current observations. This tendency is not changed even when we choose a larger $\Delta \chi^2$ cutoff value. To understand which events contribute to this change, we divide all events into two groups: high-magnification events ($A_{\rm max}>100$) and low-magnification events ($A_{\rm max}<100$). The two panels in Figure~\ref{fig:cdf-q-2sample} tell us that most of these low-mass planets are detected in low-magnification events, which is understandable since they are more often detected via planetary caustics (Figure~\ref{fig:q-s-map}).

Within all 26 well-understood microlensing planets, 3 are claimed to be super-Jupiters around M-dwarf hosts [OGLE-2005-BLG-071Lb \citep{udalski2005,dong2009a}; MOA-2009-BLG-387Lb \citep{batista2011}; OGLE-2012-BLG-406Lb \citep{poleski2014,tsapras2014}]
. The ratio, 3 out of 26, is much higher than the estimation from either core accretion theory \citep[e.g.,][]{kennedy2008} or other exoplanet detection techniques \citep[e.g.,][]{cumming2008}. In our simulation, we find that the small fraction of super-Jupiter systems given by the Ida \& Lin core accretion model is magnified by a factor of $\sim 8$ if observed via microlensing. Therefore, this observational bias should be taken into account when comparing the frequency of massive planets around M dwarfs from microlensing observations with that from planet formation theory.

Our simulation also shows that the inclination of the lens system of multiple-planet events obeys the intrinsic distribution of orbital inclinations, and that the mass ratio between the two detected planets also agrees with the intrinsic mass ratio distribution of the planetary system.

\acknowledgments
We thank the anonymous referee for useful comments. We would like to thank Shigeru Ida and Doug Lin for providing us the data from their population synthesis models and many discussions. We also thank M.B.N. Kouwenhoven and Rainer Spurzem for providing us their computing facilities. This work has also been partly supported by the Strategic Priority Research Program ``The Emergence of Cosmological Structures'' of the Chinese Academy of Sciences Grant No. XDB09000000, and by the National Natural Science Foundation of China (NSFC) under grant numbers 11333003 (SM and WZ). Work by WZ and AG was supported by NSF grant AST 1103471.

%%%%%%%%%%%%%%%%%%%%%%%%%%
\clearpage

\clearpage

\clearpage

\clearpage

\clearpage

\clearpage
\begin{table}
\centering
\caption{Number of events within each maximum magnification range. The sensitivity to planets is defined as $(N_1+2N_2)/N_{\rm total}$, where $N_1$, $N_2$ and $N_{\rm total}$ are the number of events within this class respectively. Numbers in the brackets are the numbers of caustic crossing events.
\label{tab:Amax-number}}
\small
\begin{tabular}{cccccccc}
\tableline\tableline
$A_{\rm max}$ & Total & Single-planet & Double-planet & Planetary & Central & Resonant & Sensitivity \\
($1/u_0^\star$) & events & events ($N_1$) & events ($N_2$) & caustic & caustic & caustic & to planets \\
\tableline
$3 - 20$    & 5549 & 190 & 8 & 88(50) & 69(3) & 52(35) & 3.71\%\\
$20 - 50$   & 727 & 61 & 3 & 14(13) & 37(3) & 18(14) & 9.22\%\\
$50 - 100$  & 203 & 13 & 3 & 4(4) & 10(1) & 5(4) & 9.36\%\\
$100 - 200$ & 102 & 8 & 1 & 1(1) & 7(4) & 2(2) & 9.80\%\\
\tableline
$200 - 300$ & 45 & 2 & 1 & & & & \\
$300 - 400$ & 19 & 1 & 0 & 0 & 5(3) & 1(1) & 5.50\%\\
$>400$      & 45 & 1 & 0 & & & & \\
\tableline\tableline
\end{tabular}
\end{table}

\begin{table}
\centering
\caption{Detection efficiency for planets located between $0.5R_{\rm E}$ and $2R_{\rm E}$. 
\label{tab:efficiency-mp}}
\begin{tabular}{cccc}
\tableline\tableline
$M_{\rm planet}$ & Total planets & Detected planets & Detectability \\
\tableline
$M_{\rm Mars} - M_\oplus$ & 16734 & 22 & 0.132\% \\
$M_\oplus - M_{\rm Neptune}$ & 2383 & 89 & 3.73\% \\
$M_{\rm Neptune} - M_{\rm Jupiter}$ & 183 & 80 & 43.7\% \\
$> M_{\rm Jupiter}$ & 120 & 94 & 78.3\% \\
\tableline\tableline
\end{tabular}
\end{table}

\clearpage

\clearpage

\clearpage

\clearpage

\clearpage
\appendix
\begin{sidewaystable}
  \begin{center}
  \caption{A list of all published microlensing planets and how we classify them. Planets are sorted by their mass ratio; the lower two single-solid lines show what mass ratio a Jupiter and a Neptune would have if they are around a $0.3M_\odot$ star.
  \label{tab:all-planets}}
  \small
  \begin{tabular}{lllllll}
  \tableline\tableline
  Name & $A_{\rm max}$ & $q\ (10^{-4})$ & \tabincell{c}{Caustic \\ type(s)} & \tabincell{c}{Caustic\\crossing?} & References & Comment \\
  \tableline
  \tabincell{l}{OGLE-2009-BLG-151b/\\MOA-2009-232b} & 5  & 4190 & R & Yes & \citet{choi2013} & \tabincell{l}{A brown dwarf, but listed\\as planet at \url{http://exoplanet.eu}} \\
                                OGLE-2011-BLG-0420b & 40 & 3770 & C & Yes & \citet{choi2013} & \tabincell{l}{A brown dwarf, but listed\\as planet at \url{http://exoplanet.eu}} \\
  OGLE-2012-BLG-358Lb & 10 & 800 & P & Yes & \citet{han2013b} & The host star has mass $0.02M_\odot$ \\
  MOA-2011-BLG-322Lb & 21 & 280 & C & No & \citet{shvartzvald2013} & \\
  MOA-2009-BLG-387Lb & 11 & 132 & R & Yes & \citet{batista2011} & \\
  OGLE-2005-BLG-071Lb & 42 & 71 & C & No & \citet{udalski2005} & \\
  MOA-2008-BLG-379Lb & 167 & 68.5 & R & Yes & \citet{suzuki2014} & \\
  OGLE-2012-BLG-406Lb & 2 & 59.2 & P & Yes & \citet{tsapras2014} & \\
  MOA-2011-BLG-293Lb & 286 & 53 & C & Yes & \citet{yee2012} & \\
          MOA-bin-1b & 1.1   & 49 & P & Yes & \citet{bennett2012} & \tabincell{l}{The planet has a large separation\\from the star}\\
  OGLE-2003-BLG-235Lb & 8 & 39 & R & Yes & \citet{bond2004} & \\
  \tableline
  MOA-2007-BLG-400Lb & 628 & 25 & C & Yes & \citet{dong2009b} & Same for close/wide solutions \\
  MOA-2010-BLG-477Lb & 294 & 21.81 & R & Yes & \citet{bachelet2012} & \\
  OGLE-2011-BLG-251Lb & 18 & 19.2 & C & No & \citet{kains2013} & Four solutions, D is favored\\
  OGLE-2006-BLG-109Lb & 289 & 13.5 & R & Yes & \citet{gaudi2008} & \\
  OGLE-2012-BLG-0026Lc & 109 & 7.84 & R & Yes & \citet{han2013a} & Four solutions, D is favored \\
  OGLE-2006-BLG-109Lc & 289 & 4.86 & C & Yes & \citet{gaudi2008} & \\
   MOA-2011-BLG-262Lb & 80 & 4.7 & C & Yes & \citet{bennett2014} & \tabincell{l}{An alternate model leads to\\a host mass of $\sim 4M_{\rm J}$}\\
  MOA-2009-BLG-319Lb & 167 & 3.95 & R & Yes & \citet{miyake2011} & \\
  MOA-2008-BLG-310Lb & 400 & 3.3 & C & Yes & \citet{janczak2010} & \\
  MOA-2010-BLG-328Lb & 14 & 2.6 & P & Yes & \citet{furusawa2013} & \\
  OGLE-2012-BLG-0026Lb & 109 &1.30 & C & Yes & \citet{han2013a} & Four solutions, D is favored \\
  \tableline
  OGLE-2007-BLG-368Lb & 13 & 0.95 & P & Yes & \citet{sumi2010} & \\
  OGLE-2005-BLG-169Lb & 800 & 0.9 & R & Yes & \citet{gould2006} & \\
  OGLE-2005-BLG-390Lb & 3 & 0.76 & P & Yes & \citet{beaulieu2006} & \\
  MOA-2009-BLG-266Lb & 8   & 0.563 & P & Yes & \citet{muraki2011} & \\
  MOA-2007-BLG-192Lb & $\sim270$ & -- & -- & -- & \citet{bennett2008} & \tabincell{l}{Too few data points to\\constrain the planet} \\
  \tableline\tableline
  \end{tabular}
  \end{center}
\end{sidewaystable}

\end{CJK*}
\end{document}